\documentclass[superscriptaddress, reprint, twocolumn%
 reprint,
 amsmath,amssymb,
 aps,
prb,
]{revtex4-2}

\usepackage{graphicx}
\usepackage{dcolumn}
\usepackage{bm}
\usepackage{textgreek}

\begin{document}

\preprint{APS/123-QED}

\title{Structural and magnetic properties of \textbeta-Li$_{2}$IrO$_{3}$ after grazing-angle focused ion beam thinning}

\author{Nelson Hua}
\email{nelson.hua@psi.ch}
\affiliation{Laboratory for X-Ray Nanoscience and Technologies, Paul Scherrer Institut, CH-5232 Villigen PSI, Switzerland}%
\affiliation{Quantum Criticality and Dynamics Group, Paul Scherrer Institut, CH-5232 Villigen PSI, Switzerland}%

\author{Franziska Breitner}
\affiliation{Experimental Physics VI, Center for Electronic Correlations and Magnetism, University of Augsburg, 86135 Augsburg, Germany}
\author{Anton Jesche}
\affiliation{Experimental Physics VI, Center for Electronic Correlations and Magnetism, University of Augsburg, 86135 Augsburg, Germany}
\author{Shih-Wen Huang}
\affiliation{Swiss Light Source, Paul Scherrer Institut, CH-5232 Villigen PSI, Switzerland}
\author{Christian R\"{u}egg}%
\affiliation{Quantum Criticality and Dynamics Group, Paul Scherrer Institut, CH-5232 Villigen PSI, Switzerland}%
\affiliation{Institute of Physics, Ecole Polytechnique F\'{e}d\'{e}rale de Lausanne (EPFL), CH-1015 Lausanne, Switzerland}
\affiliation{Institute for Quantum Electronics, ETH Z\"{u}rich, CH-8093 H\"{o}nggerberg, Switzerland}
\affiliation{Department of Quantum Matter Physics, University of Geneva, CH-1211 Geneva, Switzerland}
\author{Philipp Gegenwart}
\email{philipp.gegenwart@physik.uni-augsburg.de}
\affiliation{Experimental Physics VI, Center for Electronic Correlations and Magnetism, University of Augsburg, 86135 Augsburg, Germany}
\date{\today}

\begin{abstract}
Manipulating the size and orientation of quantum materials is often used to tune emergent phenomena, but precise control of these parameters is also necessary from an experimental point of view. Various synthesis techniques already exist, such as epitaxial thin film growth and chemical etching, that are capable of producing specific sample dimensions with high precision. However, certain materials exist as single crystals that are often difficult to manipulate, thereby limiting their studies to a certain subset of experimental techniques. One particular class of these materials are the lithium and sodium iridates that are promising candidates for hosting a Kitaev quantum spin liquid state. Here we  present a controlled method of using a focused ion beam at grazing incidence to reduce the size of a $\beta$-Li$_{2}$IrO$_{3}$ single crystal to a thickness of 1 $\mu m$. Subsequent x-ray diffraction measurements show the lattice remains intact, albeit with a larger mosaic spread. The integrity of the magnetic order is also preserved as the temperature dependent magnetic diffraction peak follows the same trend as its bulk counterpart with a transition temperature at \emph{T$_{N}$} = 37.5 K. Our study demonstrates a technique that opens up the possibility of nonequilibrium experiments where submicron thin samples are often essential.
\end{abstract}

\maketitle


\section{Introduction}

Enhancing certain order parameters and material properties by reducing dimensionality is a common avenue in engineering quantum materials \cite{Giustino2020, Ahn2021, Lei2017}. For example, one can drastically increase the superconducting temperature of FeSe from \emph{T$_{c}$} = 8 K to 65 K by going from a bulk crystal to a single monolayer grown on SrTiO$_{3}$ \cite{Wang2012,He2013,Kitamura2022}. In one dimension, a possible path to the realization of the topological Kondo effect has been proposed by intertwining nanowires nearby a superconducting island \cite{Beri2012,Altland2013}. Even extending to `zero' dimensional objects can give rise to peculiar properties. Confinement effects that lead to enhanced optoelectronic properties in semiconducting materials can be achieved as quantum dots where the material is reduced to a few nanometers in all directions \cite{Veldhorst2015,Pelayo2021}. These materials can be further modified with intentional or consequential defects such as strain or impurities. Nanoislands of L$_{0.7}$S$_{0.3}$MnO$_{3}$ have been synthesized from electron beam lithography with the aid of Ar$^{+}$ ion implantation that modifies the strain state, consequently enhancing the flux closure domains \cite{Takamura2006}. Introducing defects such as screw dislocations can also improve the performance of lithium-rich battery nanoparticles  \cite{Ulvestad2015, Singer2018}. The numerous existing synthesis routes has even opened up the idea of using artificial intelligence to find optimized ways to tune material properties \cite{Stanev2021,Da2023}.

On the other hand, understanding intrinsic material properties from a fundamental physics point of view often requires stringent conditions to accurately deduce the origins of emergent phenomena where the flexibility of engineering is no longer suitable. This requires the use of several complementary experimental techniques that would ideally use the same sample for all measurements, which is rarely the case. For example, we want to ensure the thin lamella sample used for transmission electron microscopy (TEM) encapsulates the same properties as the bulk crystal used for neutron scattering. This is difficult to carry out if unwanted dopants or defects were introduced in creating the lamella from the bulk sample. This also implies that thin films and bulk crystals can not always be accurately compared due to substrate-induced strain \cite{Liu2014, Liu2016}, and samples that have been processed with chemical etching may show different lattice constants due to implantation of another chemical species \cite{Takamura2006}. Furthermore, limitations such as a large lattice mismatch between the sample and substrate preclude the option of epitaxial film growth for many materials. Even when the same sample is used across different measurements, certain surface sensitive techniques may not capture the properties that are characteristic of the bulk. This is particularly noted in topological insulators where the metallic surface states do not reflect the internal insulating properties \cite{Moore2010}. Therefore, consideration of sample integrity across complementary experimental techniques is essential in studying inherent material properties. Here we present a minimally invasive use of a focused ion beam (FIB) to reduce the size of a $\beta$-Li$_{2}$IrO$_{3}$ crystal. The FIB-processed crystal is reduced to a thickness of 1 $\mu m$ and maintains its structural and magnetic properties, opening more investigative routes with experimental techniques that were previously inaccessible. 

$\beta$-Li$_{2}$IrO$_{3}$ belongs to the class of Kitaev materials, meaning it displays bond-dependent anistropic spin exchange, though additional exchanges lead to deviations from pure Kitaev physics \cite{Takayama2015, Biffin2014, Ruiz2017, Freund2016, Tsirlin2022}. The $\beta$ polymorph with a hyperhoneycomb lattice of Ir moments exhibits an incommensurate magnetic order below \emph{T$_{N}$} ~= 37 - 38 K. \cite{Takayama2015, Tsirlin2022}. Compared to other prominent Kitaev materials such as $\alpha$-RuCl$_{3}$ \cite{Sears2020} or honeycomb iridates A$_{2}$IrO$_{3}$ (A=Na, Li) \cite{Tsirlin2022} that order below 15 K, the elevated N\'eel temperature allows easier access to the  ground state by dissipative pump-probe experiments. Since its relatively recent synthesis, $\beta$-Li$_{2}$IrO$_{3}$ has been studied with x-rays, neutrons, and muons under high pressures and high magnetic fields \cite{Takayama2015, Biffin2014, Ruiz2017, Majumder2018, Majumder2019}, but only a few time-resolved experiments have looked into the dynamics and excitations in this material \cite{Glamazda2016, Choi2020}. Techniques such as ultrafast electron diffraction or time-resolved x-ray scattering can reveal dynamics of the structural and magnetic order parameters that can indirectly reveal bond-dependent interaction strengths of the correlated state \cite{Rademaker2019, Braganca2021}. However, a significant limitation is the thickness of single crystals that is often incompatible with time-resolved scattering techniques that rely on submicron thin samples. Opening these time-resolved experimental paths requires reducing these crystal thicknesses in a controlled manner.

\section{Sample and Technique}

\begin{figure} [h]
	\centering
	\includegraphics[width=0.5\textwidth]{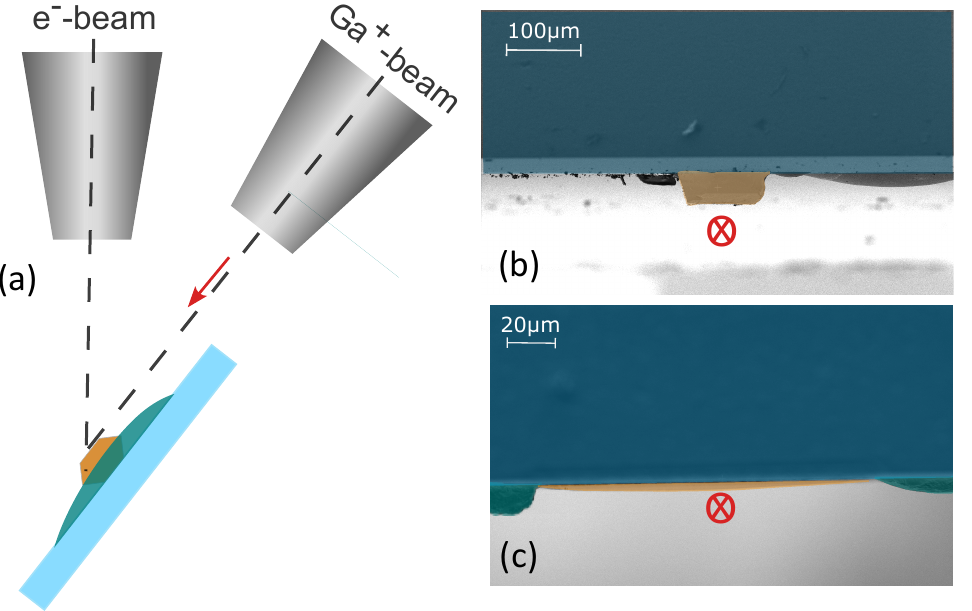}
	\caption{\label{FIB}{\bf(a)} The crystal (orange) was glued onto a substrate (light blue) using epoxy glue (teal), mounted on a pre-tilted sample holder and aligned with the surface parallel to the ion beam. Side view of the {\bf(b)} unprocessed sample and {\bf(c)} thinned sample along the ion beam (indicated by the direction of the red arrow in (a), which is also shown in red in (b) and (c) perpendicular to the picture plane).}
\end{figure}

For layered materials, thin flakes can often be easily prepared by exfoliation until the desired thickness is reached. However, this does not work for $\beta$-Li$_2$IrO$_3$ as the material does not cleave easily due to its three-dimensional structure. Another common possibility is to use a focused ion beam (FIB) for cutting a thin lamella \cite{Moll2018}. We attempted this standard milling process for $\beta$-Li$_2$IrO$_3$ with the FIB-SEM Crossbeam 500 (Zeiss), but this approach proved infeasible due to internal cracks uncovered during cutting or lamella breakage during thinning or transfer. Therefore, a different approach was required where we used the FIB-SEM to thin down the crystals by small-angle Ga-beam bombardment. A suitable crystal of sufficiently large size with distinct surfaces to assign the crystal axes by eye was chosen. After one of the surfaces perpendicular to the $c$-axis was polished, the crystal was glued onto an STO substrate (5 x 10\,mm) with the polished surface facing down using two component epoxy (Araldite Rapid) and cured at 100$^\circ$C for 1\,h. The crystal was mounted 1\,mm away from any edge to accommodate alignment requirements for future experiments. Before mounting the sample onto a pre-tilted sample holder, the sample was polished down further in order to reduce the required time of Ga beam operation. We subsequently oriented the sample in the sample chamber such that the ion beam cut parallel to the sample surface, as can be seen in Fig. \ref{FIB}(a).

\begin{figure}[h]
	\centering
	\includegraphics[width=0.45\textwidth]{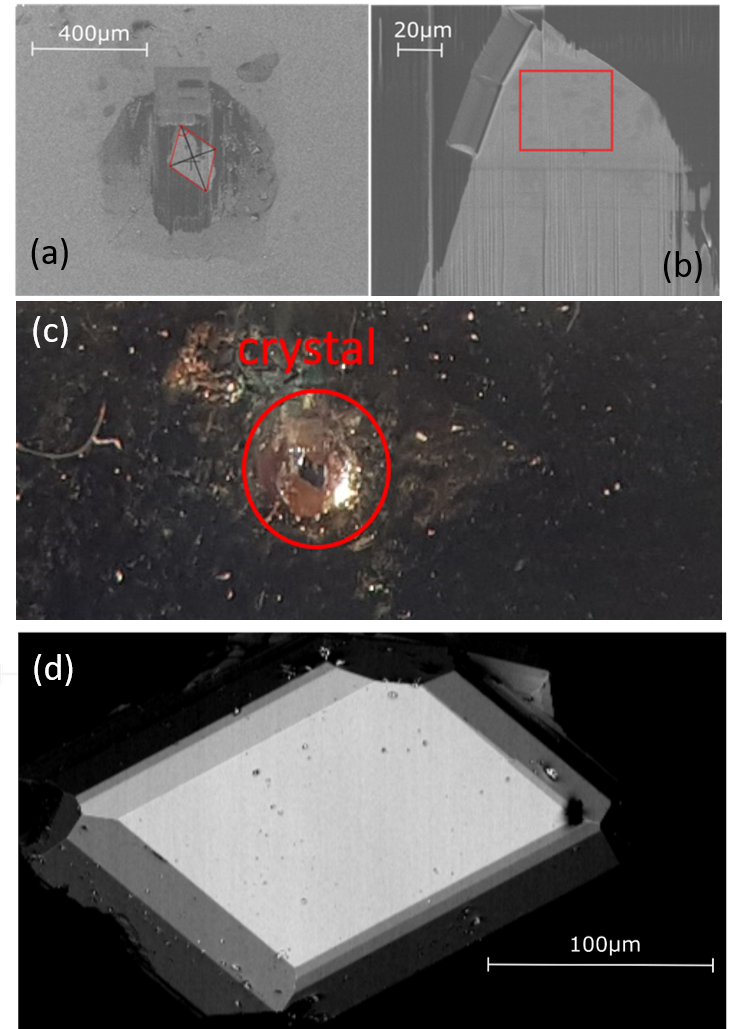}
	\caption{{\bf(a)-(c)} A FIB cut sample and {\bf(d)} bulk crystal were prepared. The red rectangle in (b) marks the measurement area while (c) shows the FIB processed sample on a gold covered STO substrate used for the x-ray diffraction measurements.} 
	\label{Probe} 
\end{figure}

Using first a voltage of 30\,kV and high probe currents up to 65\,nA, the sample was thinned, as shown in Fig. \ref{FIB}(b)-(c). In the latter stages, progressively smaller currents were applied before finally removing possible Ga implantations by performing low energy milling. In this way, an area of approximately 50 x 50\,$\mu m$ was thinned down to a thickness of about 1\,$\mu m$. Attempts to further thin the sample resulted in the onset of peeling at the edges of the crystal. Thus, no further thinning of the sample depicted in Fig. \ref{Probe} (a)-(c) was performed. For comparison, an untreated, bulk crystal with a surface area of about 100 x 150\,$\mu m$ and a thickness of 50 $\mu m$, as depicted in Fig. \ref{Probe}(d), was mounted onto an STO substrate with the $c$-axis orientated perpendicular to the substrate surface.

\section{Experiment and Results}

The x-ray diffraction experiment on the bulk and FIB-processed $\beta$-Li$_{2}$IrO$_{3}$ crystals was carried out at the Materials Science Surface Diffraction (X04SA) beamline of the Swiss Light Source \cite{MS}. The x-ray energy was tuned to 11.115 keV to measure the (004) and (135) lattice peaks at room temperature while 11.215 keV was used to measure the incommensurate (-0.584, 0, 16) magnetic peak at the Ir \emph{L$_{3}$}-edge resonance. We followed the evolution of the magnetic peak in the FIB-processed crystal between 15.0 - 37.5 K that can only be detected at resonance below \emph{T$_{N}$}. Both crystals were aligned with the  $c$-axis out of plane and the diffraction signals were recorded on a 0.5M Eiger detector with 75 $\mu$m pixel sizes. The experimental geometry follows a standard (2+3) surface diffractometer where the 2D detector is fixed approximately 1 meter from the sample and moves along a spherical surface in the 2$\theta$ and $\delta$ directions.

\begin{figure}
	\includegraphics[width=0.45\textwidth]{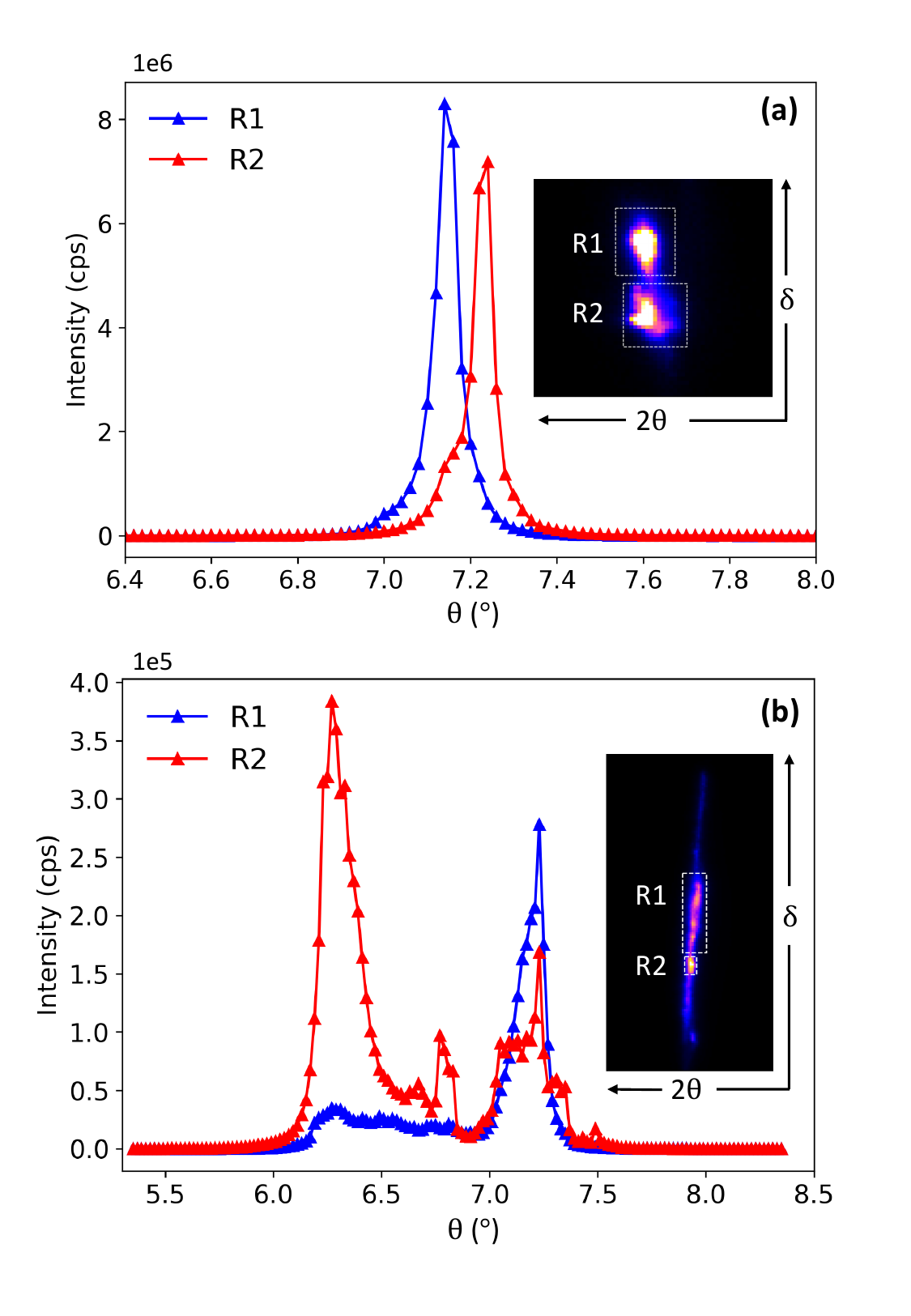}
	\caption{{\bf(a)} Rocking curve of the bulk crystal (004) lattice peak that is split into two domains designated as R1 and R2. (Inset) The two peaks appear at the same 2$\theta$ value of the detector, but at a different $\delta$ value. {\bf(b)} The (004) lattice peak of the FIB-processed crystal shows multiple domains and the corresponding summed detector image is shown in the inset. Two principal regions are designated as R1 and R2.}
	\label{Fig. 2}
\end{figure}

\begin{figure}[h]
	\includegraphics[scale=0.45]{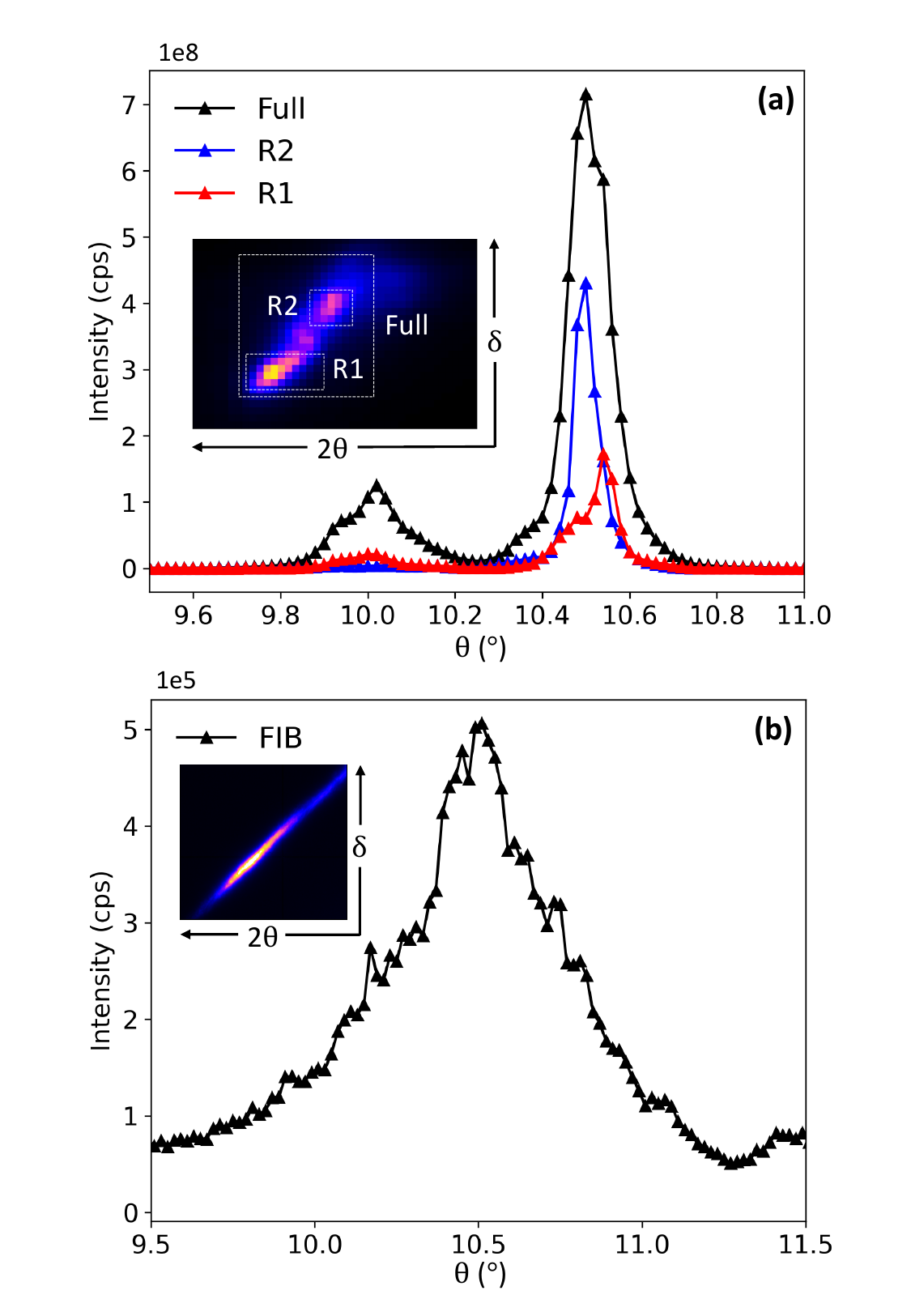}
	\caption{\label{Fig3}  {\bf(a)} Rocking curve of the bulk crystal (135) lattice peak and the corresponding detector regions of interest, R1 and R2, shown in the inset.  {\bf(b)} The (135) lattice peak of the FIB-processed crystal that exhibits a large mosaic spread.}
\end{figure}

The (004) diffraction peak directly probes the out-of-plane lattice structure, including the crystallographic $c$-axis lattice constant. The measured lattice constants from the (004) peak of the bulk and the FIB-processed samples were 1.783 \AA\, and 1.787 \AA, respectively. These are in line with previously reported values of 1.779 \AA\, \cite{Biffin2014} and 1.786 \AA\, \cite{Ruiz2017}. As seen in Fig. \ref{Fig. 2}(a), the (004) lattice peak is split on the detector along the $\delta$ direction and with a slight offset in the incidence angle $\theta$. This shows that our bulk crystal consists of two principal domains that are slightly tilted from each other. However, their 2$\theta$ values are the same, indicating both domains have the same lattice constant. Fig. \ref{Fig. 2}(b) shows the (004) lattice peak profile of the FIB-processed sample that is broken into several domains. The rocking curves of two regions of interest on the detector, designated as R1 and R2 in the inset, are shown where the diffraction peak is spread along approximately the same  2$\theta$ value. Therefore, these domains share the same lattice constant, but the larger spread in the sample $\theta$ and detector $\delta$ directions directly translates to a larger mosaic spread. The (135) lattice peak additionally probes the in-plane lattice integrity where the rocking curves for the bulk and FIB-processed crystals are shown in Fig. \ref{Fig3}(a) and (b), respectively. Similar to the (004) lattice peak, the (135) lattice peak of the bulk crystal shows up at two sample $\theta$ angles and two locations on the detector, indicating two principal domains. On the other hand, the FIB-processed crystal shows primarily one wide, continuous peak in both the sample $\theta$ angle and on the detector, indicative of a mosaic spread with small domain sizes.

\begin{figure}[h]
	\includegraphics[width=.45\textwidth]{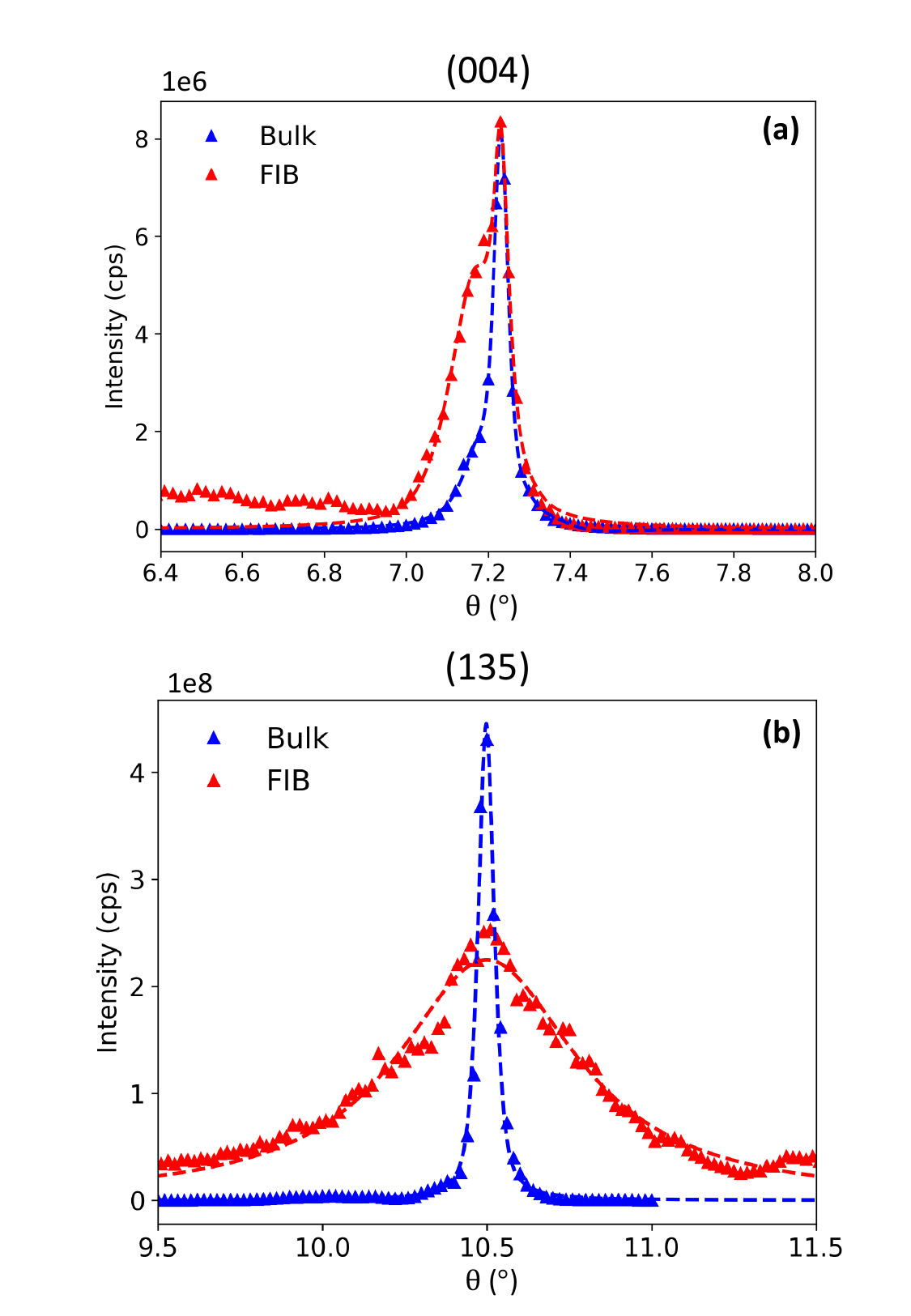}
	\caption{\label{Fig4} {\bf(a)} The (004) lattice peak of the bulk crystal R2 domain superimposed on the corresponding R1 domain of the FIB-processed crystal (magnified by 30x). A double pseudo-Voigt function was used as a fit to take into account the shoulder. {\bf(b)} The (135) lattice peak of the bulk crystal R1 domain superimposed on the corresponding FIB-processed crystal lattice peak (magnified by 500x). Both peaks were fitted with a single pseudo-Voigt function.}
\end{figure}

To compare the effects of our technique quantitatively, both the (004) and (135) lattice peaks of the bulk and FIB-processed crystals were fitted with single and double pseudo-Voigt functions. Fig. \ref{Fig4}(a) shows the (004) lattice peak of the bulk crystal R2 domain superimposed on the FIB-processed crystal R1 domain. The effective full width half maximums (FWHM) from the fits are 0.048$^\circ$ and 0.126$^\circ$, resulting in a correlation length increase of 2.6 times larger for the bulk crystal. Fig. \ref{Fig4}(b) shows the (135) lattice peaks of the bulk crystal R1 domain and the  FIB-processed crystal. The FWHM of the bulk crystal diffraction peak is 0.068$^\circ$ whereas the FWHM of the FIB-processed crystal diffraction peak is 0.600$^\circ$, which is a 9.4 times larger correlation length for the bulk crystal. The much larger change in correlation length of the (135) lattice peak suggests that our FIB technique induces a more significant change to the in-plane structural integrity compared of the out-of-plane lattice structure. 

\begin{figure*}
	\includegraphics[width=1\textwidth]{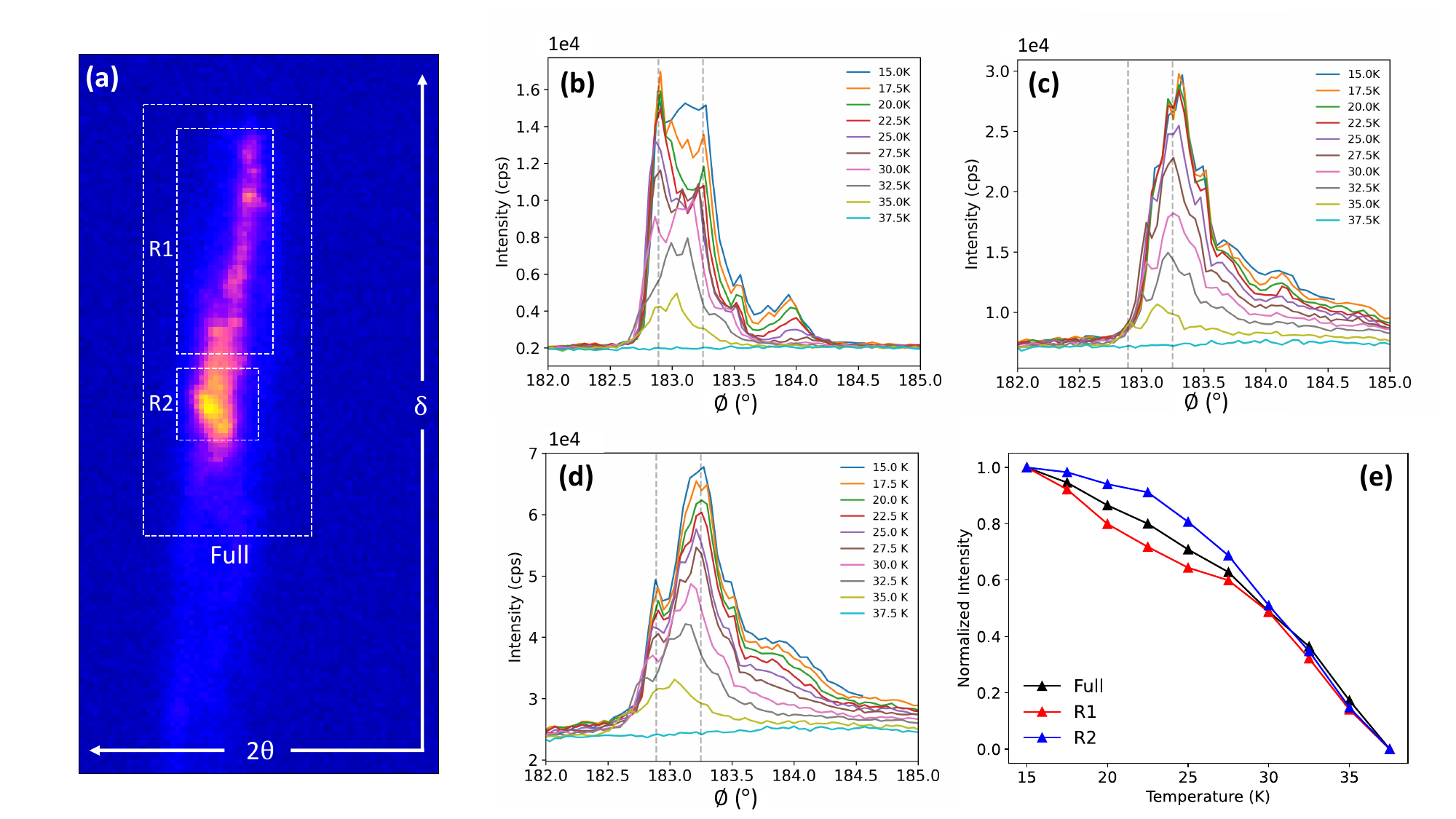}
	\caption{\label{Fig5} {\bf(a)} The summed detector image of a rocking curve on the (-0.584, 0, 16) magnetic peak at \emph{T} = 15 K. The rocking curves of the {\bf(b)} R1, {\bf(c)} R2, and the {\bf(d)} full diffraction peak  regions of interest between 15.0 - 37.5 K are shown. The vertical dotted lines are reference lines to compare notable peaks for the two regions of interest. {\bf(e)} The normalized integrated intensity of the  regions of interest as a function of temperature.}   
\end{figure*}

Finally, the magnetic structure of the FIB-processed crystal was studied via the incommensurate (-0.584, 0, 16) resonant magnetic peak. The summed detector image of a rocking curve at \emph{T} = 15 K is shown in Fig. \ref{Fig5}(a). Here the diffraction peak is again broadened along the same 2$\theta$ value, which is in line with the (004) lattice structure as the diffraction geometry of the magnetic structure is almost specular. This suggests that the boundaries of the magnetic domain structure are defined by the lattice domains. The rocking curves of the incommensurate peak was performed by rotating $\phi$, the azimuthal angle about the $c$-axis lattice direction, to maintain a constant penetration depth. The rocking curves for the regions of interest, defined as R1, R2 and the full diffraction peak in Fig. \ref{Fig5}(a), are shown in Figs. \ref{Fig5}(b)-(d) for the temperature range between 15.0 - 37.5 K. The normalized integrated intensity from these rocking curves are shown in Fig. \ref{Fig5}(e) where the peak disappears completely at \emph{T} = 37.5 K. This aligns well with previously measured transition temperatures for a single crystal at \emph{T$_{N}$} = 37 - 38 K \cite{Tsirlin2022, Ruiz2017}. 

\section{Conclusion}
We have demonstrated a controlled, FIB-based technique that is capable of reducing the thickness of $\beta$-Li$_{2}$IrO$_{3}$ crystals. Since this material neither easily cleaves nor allows FIB-assisted thin lamella cutting, the new technique provides a unique possibility to obtain  $\beta$-Li$_{2}$IrO$_{3}$ crystal thicknesses below $\sim$ 1 $ \mu m$. As expected, the processed crystal exhibits a larger mosaic spread resulting in smaller domains not present in the original bulk crystal. However, the electronic structure of the material, namely the magnetically ordered state, is still preserved and follows the same temperature-dependent trend as its bulk counterpart. The $c$-axis lattice constant of the FIB-processed crystal is also in line with previously measured values from bulk samples, suggesting that the FIB process did not introduce dopants that would have significantly strained the sample. Currently the most significant limitation of this technique is the onset of peeling as the sample approaches nanometer thicknesses, but further improvements to this technique in the future can result in crystal sizes with a thickness of several hundred nanometers. Extending this to other materials where neither cleaving nor thin lamella preparation is possible will open the possibility of time-resolved experiments such as ultrafast electron diffraction or pump-probe x-ray diffraction measurements.

\section{Acknowledgements}
This project has received funding from the European Union's Horizon 2020 research and innovation programme under the Marie Sklodowska-Curie grant agreement No 884104 (PSI-FELLOW-III-3i). This work was funded by the Deutsche Forschungsgemeinschaft (DFG, German Research Foundation) -- TRR 360 -- 492547816. We also acknowledge the Paul Scherrer Institut, Villigen, Switzerland for provision of synchrotron radiation beamtime at beamline X04SA of the SLS.

\bibliography{Manuscript}

\end{document}